\documentclass[aps,preprint,prd,nofootinbib]{revtex4}
\usepackage{graphicx}
\usepackage{psfrag}

\begin{document}

\title{The Strong Decays of $X(3940)$ and $X(4160)$ }
\author{Zhi-Hui Wang$^{[1],[2]}$\footnote{2013086@nun.edu.cn}, Yi Zhang$^{[1]}$, Li-Bo Jiang$^{[3]}$, Tian-hong Wang$^{[2]}$, Yue Jiang$^{[2]}$,
Guo-Li Wang$^{[2]}$\\}
\address{$^1$School of Electrical $\&$ Information Engineering, Beifang University of Nationalities, Yinchuan, 750021, People’s Republic of
China\\
$^2$Department of Physics, Harbin Institute of
Technology, Harbin, 150001, People’s Republic of
China\\
$^3$Department of physics and astronomy, University of Pittsburgh, PA 15260, USA}

 \baselineskip=20pt

\begin{abstract}
The new mesons $X(3940)$ and $X(4160)$ have been found by Belle Collaboration in the processes $e^+e^-\to J/\psi D^{(*)}\bar D^{(*)}$.
Considering $X(3940)$ and $X(4160)$ as $\eta_c(3S)$ and $\eta_c(4S)$ states,
the two-body open charm OZI-allowed strong decay of $\eta_c(3S)$ and $\eta_c(4S)$ are studied by the improved Bethe-Salpeter
method combine with the $^3P_0$ model.
The strong decay width of $\eta_c(3S)$ is $\Gamma_{\eta_c(3S)}=(33.5^{+18.4}_{-15.3})$ MeV,
which is closed to the result of $X(3940)$,
therefore, $\eta_c(3S)$ is a good candidate of $X(3940)$.
The strong decay width of $\eta_c(4S)$ is
$\Gamma_{\eta_c(4S)}=(69.9^{+22.4}_{-21.1})$ MeV,
considering the errors of the results,
it's closed to the lower limit of $X(4160)$.
But the ratio of the decay width $\frac{\Gamma(D\bar D^*)}{\Gamma (D^*\bar D^*)}$ of $\eta_c(4S)$ is larger than
the experimental data of $X(4160)$.
According to the above analysis,
$\eta_c(4S)$ is not the candidate of $X(4160)$,
and more investigations of $X(4160)$ is needed.
 \vspace*{0.5cm}

\noindent {\bf Keywords:} $X(3940)$; $X(4160)$;
  Strong Decay; Improved Bethe-Salpeter Method.

\end{abstract}

\maketitle
\section{Introduction}

In the past few years, many more new charmonium-like states, so-called $XYZ$ states,
have been observed by the Belle, BABAR and BESIII Collaborations~\cite{PDG}.
The discovery of these states not only enriched the spectroscopy of charmonium-like states
but also provided us an opportunity to research the properties of charmonium-like states.
For example,
the $X(3940)$ state was observed from the inclusive process $e^+e^-\to J/\psi X(3940)$
and had the decay mode $X(3940)\to D^*\bar D$ by the Belle Collaboration at a mass of $(3943\pm6\pm6)$ MeV.
The decay width of these state was less than 52 MeV at the $90\%$ C.L. which has taken into the systematics~\cite{3940}.
Later Belle Collaboration confirmed the observation of $X(3940)$ with a significance
of $5.7\sigma$, they got the mass and width of $X(3940)$ were $M=(3942^{+7}_{-6}\pm6)$ MeV,
$\Gamma=(37^{+26}_{-15}\pm8)$ MeV.
At the same time, they also observed a new charmonium-like state $X(4160)$
in the process $e^+e^-\to J/\psi D^*\bar D^*$,
the mass and width of $X(4160)$ were
$M=(4156^{+25}_{-20}\pm15)$ MeV, $\Gamma=(139^{+111}_{-61}\pm21)$ MeV~\cite{39404160}.

The observations of these $XYZ$ states inspire many interests about their physical
natures.
There are already many theoretical approaches which have been used
to study the properties of these $XYZ$ states~\cite{liuxiang1,zhushilin1,barguta,zhao1,th1,th2,th3,th4,th5,th6,th7,th8,39401,41601,liu2,Zhurl,Yan1}.
In this paper, we mainly discuss the properties and decays of two $X$ states: $X(3940)$ and $X(4160)$.
Ref.~\cite{liuxiang1} and Ref.~\cite{zhushilin1} have assigned that
the $C$ parity of $X(3940)$ and $X(4160)$ should be even, $C=+$.
Assuming $X(3940)$ as $3^1S_0$ or one of $2^3P_J$ charmonium-like states,
Ref.~\cite{barguta} studied $e^+e^-\to J/\psi X(3940)$ process by the light-cone formalism,
and they considered that $X(3940)$ is $3^1S_0$($\eta_c(3S)$).
Ref.~\cite{zhao1,th1} investigated the properties of $X(3940)$ and $X(4160)$ which were $\eta_c(3S)$ and $\chi_{c0}(3P)$, respectively.
Ref.~\cite{th6} calculated the strong decays of $\eta_c(nS)$,
they found that the explanation of $X(3940)$
as $\eta_c(3S)$ is possible and the assignment of $X(4160)$ as $\eta_c(4S)$ can not
be excluded.
In Ref.~\cite{th7}, the authors studied vector-vector interaction of $X(4160)$
which was basically a $D_s^*\bar D_s^*$ molecular state with $J^{PC}=2^{++}$.
Ref.~\cite{39401} had studied the inclusive production of $X(3940)$
in the decay of ground bottomonium state $\eta_b$ by the NRQCD factorization formula,
and they also considered $X(3940)$ as the excited $\eta_c(3S)$ state.
Ref.~\cite{41601} calculated the strong decay of $X(4160)$ which was assumed as $\chi_{c0}(3P)$,
$\chi_{c1}(3P)$, $\eta_{c2}(2D)$ or $\eta_c(4S)$ by the $^3P_0$ model.
In Ref.~\cite{liu2}, they also explored the properties of $X(3940)$ and $X(4160)$
as the $\eta_c(3S)$ and $\eta_c(4S)$, respectively.
But their results suggested that $X(3940)$ as $\eta_c(3S)$ was established and
the explanation of $X(4160)$ to
be $\eta_c(4S)$ is fully excluded.
Using the NRQCD factorization approach, Ref.~\cite{Zhurl} calculated the branching fractions of
$\Upsilon(nS)\to J/\psi+X$ with $X=X(3940)$ or $X=X(4160)$,
they thought that the $X(3940)$ and $X(4160)$ can be explained as $3^1S_0$ and $4^1S_0$ charmonium-like states, respectively.
Up to now, it is very difficult to confirm the constructions of $X(3940)$ and $X(4160)$,
because lack of the enough experimental data.
Many more theoretical prediction and experimental data are needed for $X(3940)$ and $X(4160)$.

The mesons can be described by the B-S equation.
Ref.~\cite{robert1} took the B-S equation to describe the light mesons $\pi$ and $K$,
then they calculated the mass and decay constant of $\pi$ by the B-S amplitudes~\cite{robert2},
they also studied the weak decays~\cite{robert3} and the strong decays~\cite{robert4} combine with the Dyson-Schwinger
equation.

We will use the B-S equation to study the properties of heavy mesons.
In Ref.\cite{mass1}, we had calculate the Spectrum of heavy quarkonia the improved Bethe-Salpeter(B-S) method,
for the the charmonium state with the quantum numbers $J^{PC} =0^{-+}$,
the mass of $3^1S_0$ ($\eta_c(3S)$) is $M$=3948.8 MeV which is closed to the mass of $X(3940)$ with error,
the mass of $4^1S_0$ ($\eta_c(4S)$) was $M$=4224.6 MeV which was larger than the center mass of $X(4160)$ about 70 MeV.
In this paper, to check if the $X(4160)$ is the charmonium $\eta_c(4S)$, we calculate the strong decay of $\eta_c(4S)$,
but assign the mass of $\eta_c(4S)$ as 4156 MeV by varying the parameter $V_0$ in interaction potential,
where in potential model the parameter $V_0$ is added to move the theoretical mass spectra parallel to match the experimental data.

Using the the improved B-S method,
we calculated the weak decay of $B_c$ to $\eta_c(1S)$ and $\eta_c(2S)$~\cite{etac1},
and the weak decay of $B_c$ to $\eta_c(3S)$ and $\eta_c(4S)$~\cite{3940-wang}.
There is nobody to calculate $B_c$ to $\eta_c(4S)$ ,
but the results of $B_c$ to $\eta_c(1S)$, $\eta_c(2S)$, $\eta_c(3S)$ were close to the other theoretical results.
We also studied the properties of some $XYZ$ states,
such as radiative E1 decay of $X(3872)$~\cite{th2,th3},
two-body strong decay of $Z(3930)$ which was $\chi_{c2}(2P)$ state combine with the $^3P_0$ model~\cite{th4},
the strong decay of $X(3915)$ as $\chi_{c0}(2P)$ state~\cite{YJiang},
and the strong decay of $\Upsilon$~\cite{3p06}.
All the theoretical results consist with experimental data or other theoretical results.
Because the higher excited states have larger relativistic correction than
the corresponding ground state, a relativistic model is needed in a careful study.
The improved B-S method is a relativistic model that describe bound states with definite quantum number,
the corresponding relativistic form of wavefunctions are solutions of the full Salpeter equations.
So the improved B-S method is good method to describe the properties and decays of the radial high excited states,
In this paper, we focus on the strong decays of $X(3940)$ and $X(4160)$
as radial high excited states $\eta_c(3S)$ and $\eta_c(4S)$ by the improved B-S method.

In our method, we study the natures of heavy mesons by the coupling of $L+S$ for the quark and anti-quark in mesons.
According to the $L+S$ coupling, we show the wavefunctions of the heavy mesons in term of the quantum number $J^P$
(or $J^{PC}$) which are very good to describe the equal mass systems in heavy mesons.
The quantum numbers $J^{PC}$ of $\eta_c(3S)$ and $\eta_c(4S)$ both are $0^{-+}$,
the $C$ parities are even which agree with the results of Ref.~\cite{liuxiang1} and Ref.~\cite{zhushilin1}.
The corresponding Okubo–-Zweig–-Iizuka (OZI)~\cite{OZI1,OZI2,OZI3} rule allowed two-body open charm strong decay modes are:
$0^-\to 0^-1^-$ and $0^-\to 1^-1^-$, while other strong decays in the final state are ruled
out by the kinematic possible mass region.
In order to calculate the two-body open charm strong decay,
we adopt the $^3P_0$ model which assumes that a quark-antiquark pairs is created with vacuum quantum
numbers, $J^{PC}=0^{++}$~\cite{3p01,3p02,3p03}.
The $^3P_0$ model was proposed in Ref.~\cite{3p01},
then Ref.~\cite{3p02} and Ref.~\cite{3p03} applied the $^3P_0$ model to
study the open-flavor strong decays of the light mesons.
Now, People have extended this model to study the natures of heavy-light mesons~\cite{3p0hl1,3p0hl2}
and heavy quarkonia~\cite{3p04,3p05,3p06}.
In Ref.~\cite{th4} and Ref.~\cite{3p06},
we have calculated the OZI allowed two-body strong decays of charmonium and bottomonium
in the $^3P_0$ model with the relativistic B-S wavefunctions.
The results were good according with experimental data and the other theoretical results.
Furthermore, the strong decay widths are related to the parameter $\gamma$,
but the ratio of the decay width $\frac{\Gamma(\eta_c(4S)\to D\bar{D}^*}{\Gamma(\eta_c(4S)\to D^*\bar{D}^*}$ and
$\frac{\Gamma(\eta_c(4S)\to D\bar{D}}{\Gamma(\eta_c(4S)\to D^*\bar{D}^*}$ were independent of the parameter $\gamma$,
so the results of the ratios are more reliable than the decay widths.
In this paper, we take the same method as Ref.~\cite{th4} and Ref.~\cite{3p06} to study
strong decays of $\eta_c(3S)$ and $\eta_c(4S)$ states.

The paper is organized as follows.
In Sec.~II, we introduce the instantaneous
B-S equation;
We show the relativistic wavefunctions of initial mesons and final mesons in Section.~III;
In Sec.~IV, we give the formulation of two-body open charm strong decays;
The corresponding results and conclusions are present in Sec.~V.
\section{Instantaneous Bethe-Salpeter Equation}

In this section, we briefly review the Bethe-Salpeter equation and
its instantaneous one, the Salpeter equation.

The BS equation is read as~\cite{BS}:
\begin{equation}
(\not\!{p_{1}}-m_{1})\chi(q)(\not\!{p_{2}}+m_{2})=
i\int\frac{d^{4}k}{(2\pi)^{4}}V(P,k,q)\chi(k)\;, \label{eq1}
\end{equation}
where $\chi(q)$ is the B-S wave function, $P$ is the total momentum of the meson,
$q$ is relative quantum between quark and anti-quark, $V(P,k,q)$ is the
interaction kernel between the quark and anti-quark, $p_{1},
p_{2}$ and $m_1$, $m_2$ are the momentum and mass of the quark 1 and anti-quark 2, respectively.

We divide the relative momentum $q$ into two parts,
$q_{\parallel}$ and $q_{\perp}$,
$$q^{\mu}=q^{\mu}_{\parallel}+q^{\mu}_{\perp}\;,$$
$$q^{\mu}_{\parallel}\equiv (P\cdot q/M^{2})P^{\mu}\;,\;\;\;
q^{\mu}_{\perp}\equiv q^{\mu}-q^{\mu}_{\parallel}\;.$$

Correspondingly, we have two Lorentz invariants:
\begin{center}
$q_{p}=\frac{(P\cdot q)}{M}\;, \;\;\;\;\;
q_{_T}=\sqrt{q^{2}_{p}-q^{2}}=\sqrt{-q^{2}_{\perp}}\;.$
\end{center}
When $\stackrel{\rightarrow}{P}=0$, $q_p=q_{0}$ and $q_T=|\vec q|$, respectively.

In instantaneous approach, the kernel $V(P,k,q)$ takes the simple
form~\cite{Salp}:
$$V(P,k,q) \Rightarrow V(|\vec k-\vec q|)\;.$$

Let us introduce the notations $\varphi_{p}(q^{\mu}_{\perp})$ and
$\eta(q^{\mu}_{\perp})$ for three dimensional wave function as
follows:
$$
\varphi_{p}(q^{\mu}_{\perp})\equiv i\int
\frac{dq_{p}}{2\pi}\chi(q^{\mu}_{\parallel},q^{\mu}_{\perp})\;,
$$
\begin{equation}
\eta(q^{\mu}_{\perp})\equiv\int\frac{dk_{\perp}}{(2\pi)^{3}}
V(k_{\perp},q_{\perp})\varphi_{p}(k^{\mu}_{\perp})\;. \label{eq5}
\end{equation}
Then the BS equation can be rewritten as:
\begin{equation}
\chi(q_{\parallel},q_{\perp})=S_{1}(p_{1})\eta(q_{\perp})S_{2}(p_{2})\;.
\label{eq6}
\end{equation}
The propagators of the two constituents can be decomposed as:
\begin{equation}
S_{i}(p_{i})=\frac{\Lambda^{+}_{ip}(q_{\perp})}{J(i)q_{p}
+\alpha_{i}M-\omega_{i}+i\epsilon}+
\frac{\Lambda^{-}_{ip}(q_{\perp})}{J(i)q_{p}+\alpha_{i}M+\omega_{i}-i\epsilon}\;,
\label{eq7}
\end{equation}
with
\begin{equation}
\omega_{i}=\sqrt{m_{i}^{2}+q^{2}_{_T}}\;,\;\;\;
\Lambda^{\pm}_{ip}(q_{\perp})= \frac{1}{2\omega_{ip}}\left[
\frac{\not\!{P}}{M}\omega_{i}\pm
J(i)(m_{i}+{\not\!q}_{\perp})\right]\;, \label{eq8}
\end{equation}
where $i=1, 2$ for quark and anti-quark, respectively,
 and
$J(i)=(-1)^{i+1}$.

Introducing the notations $\varphi^{\pm\pm}_{p}(q_{\perp})$ as:
\begin{equation}
\varphi^{\pm\pm}_{p}(q_{\perp})\equiv
\Lambda^{\pm}_{1p}(q_{\perp})
\frac{\not\!{P}}{M}\varphi_{p}(q_{\perp}) \frac{\not\!{P}}{M}
\Lambda^{{\pm}}_{2p}(q_{\perp})\;. \label{eq10}
\end{equation}

With contour integration over $q_{p}$ on both sides of
Eq.~(\ref{eq6}), we obtain:
$$
\varphi_{p}(q_{\perp})=\frac{
\Lambda^{+}_{1p}(q_{\perp})\eta_{p}(q_{\perp})\Lambda^{+}_{2p}(q_{\perp})}
{(M-\omega_{1}-\omega_{2})}- \frac{
\Lambda^{-}_{1p}(q_{\perp})\eta_{p}(q_{\perp})\Lambda^{-}_{2p}(q_{\perp})}
{(M+\omega_{1}+\omega_{2})}\;,
$$
and the full Salpeter equation:
$$
(M-\omega_{1}-\omega_{2})\varphi^{++}_{p}(q_{\perp})=
\Lambda^{+}_{1p}(q_{\perp})\eta_{p}(q_{\perp})\Lambda^{+}_{2p}(q_{\perp})\;,
$$
$$(M+\omega_{1}+\omega_{2})\varphi^{--}_{p}(q_{\perp})=-
\Lambda^{-}_{1p}(q_{\perp})\eta_{p}(q_{\perp})\Lambda^{-}_{2p}(q_{\perp})\;,$$
\begin{equation}
\varphi^{+-}_{p}(q_{\perp})=\varphi^{-+}_{p}(q_{\perp})=0\;.
\label{eq11}
\end{equation}

For the different $J^{PC}$ (or $J^{P}$) states, we give the general form of
wave functions. Reducing the wave functions by the last
equation of Eq.~(\ref{eq11}), then solving the first and second equations in Eq.~(\ref{eq11}) to
get the wave functions and mass spectrum. We have discussed the
solution of the Salpeter equation in detail in Ref.~\cite{w1,mass1}.

The normalization condition for BS wave function is:
\begin{equation}
\int\frac{q_{_T}^2dq_{_T}}{2{\pi}^2}Tr\left[\overline\varphi^{++}
\frac{{/}\!\!\!
{P}}{M}\varphi^{++}\frac{{/}\!\!\!{P}}{M}-\overline\varphi^{--}
\frac{{/}\!\!\! {P}}{M}\varphi^{--}\frac{{/}\!\!\!
{P}}{M}\right]=2P_{0}\;. \label{eq12}
\end{equation}

 In our model, the instantaneous interaction kernel $V$ is Cornell
potential, which is the sum of a linear scalar interaction and a vector interaction:
\begin{equation}\label{vrww}
V(r)=V_s(r)+V_0+\gamma_{_0}\otimes\gamma^0 V_v(r)= \lambda
r+V_0-\gamma_{_0}\otimes\gamma^0\frac{4}{3}\frac{\alpha_s}{r}~,
\end{equation}
 where $\lambda$ is the string constant and $\alpha_s(\vec
q)$ is the running coupling constant. In order to fit the data of
heavy quarkonia, a constant $V_0$ is often added to confine
potential. To avoid the infrared divergence $V_v({\vec q})$ at $q=0$ in the momentum space, we introduce a factor $e^{-\alpha r}$ to avoid
the divergence:
\begin{equation}
V_s(r)=\frac{\lambda}{\alpha}(1-e^{-\alpha r})~,
~~V_v(r)=-\frac{4}{3}\frac{\alpha_s}{r}e^{-\alpha r}~.
\end{equation}\label{vsvv}
 It is easy to
know that when $\alpha r\ll1$, the potential becomes to Eq.~(\ref{vrww}). In the momentum space and the C.M.S of the bound state,
the potential reads :
$$V(\vec q)=V_s(\vec q)
+\gamma_{_0}\otimes\gamma^0 V_v(\vec q)~,$$
\begin{equation}
V_s(\vec q)=-(\frac{\lambda}{\alpha}+V_0) \delta^3(\vec
q)+\frac{\lambda}{\pi^2} \frac{1}{{(\vec q}^2+{\alpha}^2)^2}~,
~~V_v(\vec q)=-\frac{2}{3{\pi}^2}\frac{\alpha_s( \vec q)}{{(\vec
q}^2+{\alpha}^2)}~,\label{eq16}
\end{equation}
where the running coupling constant $\alpha_s(\vec q)$ is :
$$\alpha_s(\vec q)=\frac{12\pi}{33-2N_f}\frac{1}
{\log (a+\frac{{\vec q}^2}{\Lambda^{2}_{QCD}})}~.$$ We introduce a small
parameter $a$ to
avoid the divergence in the denominator. The constants $\lambda$, $\alpha$, $V_0$ and
$\Lambda_{QCD}$ are the parameters that characterize the potential. $N_f=3$ for $\bar bq$ (and $\bar cq$) system.

\section{The Relativistic Wavefunctions}
In this paper, we focus on the two-body open charm strong decay
of $X(3940)$ and $X(4160)$ which are considered as $\eta_c(3S)$ $\eta_c(4S)$ states.
$\eta_c(3S)$ $\eta_c(4S)$ states have two decay modes:
$0^-\to 0^-1^-$ and $0^-\to 1^-1^-$.
So we only discuss the relativistic wavefunctions of $J^{P}$ equal to $0^{-}(^1S_0)$ and $1^-(^3S_1)$ states.

\subsection{ For pseudoscalar meson with quantum numbers
$J^{P}=0^{-}$}

The general form for the relativistic wavefunction of
pseudoscalar meson can be written as~\cite{w1}:
\begin{eqnarray}\label{aa01}
\varphi_{0^-}(\vec q)&=&\Big[f_1(\vec q){\not\!P}+f_2(\vec q)M+
f_3(\vec q)\not\!{q_\bot}+f_4(\vec q)\frac{{\not\!P}\not\!{q_\bot}}{M}\Big]\gamma_5,
\end{eqnarray}
where $M$ is the mass of the pseudoscalar meson, and
$f_i(\vec q)$ are functions of $|\vec q|^2$. Due to
the last two equations of Eq.~(\ref{eq11}):
$\varphi_{0^-}^{+-}=\varphi_{0^-}^{-+}=0$, we have:
\begin{eqnarray}\label{constrain}
f_3(\vec q)&=&\frac{f_2(\vec q)
M(-\omega_1+\omega_2)}{m_2\omega_1+m_1\omega_2},~~~
f_4(\vec q)=-\frac{f_1(\vec q)
M(\omega_1+\omega_2)}{m_2\omega_1+m_1\omega_2}.
\end{eqnarray}
where $m_1, m_2$ and
$\omega_1=\sqrt{m_1^{2}+\vec{q}^2},\omega_2=\sqrt{m_2^{2}+\vec{q}^2}$ are
the masses and the energies of
quark and anti-quark in mesons, $q_{\bot}^2=-|\vec q|^2$.

The numerical values of radial wavefunctions $f_1$, $f_2$ and
eigenvalue $M$ can be obtained by solving the first two Salpeter equations in
 Eq.~(\ref{eq11}).
 In Ref.~\cite{3940-wang}, we have plot the wavefunctions of $X(3940)$ and $X(4160)$
 which are considered as $\eta_c(3S)$ and $\eta_c(4S)$, respectively.

According to the Eq.~(\ref{eq10}) the relativistic positive wavefunction
of pseudoscalar meson in C.M.S can be written as \cite{w1}:
\begin{eqnarray}\label{0-postive}
{\varphi}^{++}_{0^-}(\vec{q})=b_1
\left[b_2+\frac{\not\!{P}}{M}+b_3\not\!{q_{\bot}}
+b_4\frac{\not\!{q_{\bot}}\not\!{P}}{M}\right]{\gamma}_5,
\end{eqnarray}
where the $b_i$s ($i=1,~2,~3,~4$) are related to the original
radial wavefunctions $f_1$, $f_2$, quark masses $m_1$, $m_2$, quark energy $w_1$, $w_2$,
and meson mass $M$:
$$b_1=\frac{M}{2}\left({f}_{1}(\vec{q})
+{f}_{2}(\vec{q})\frac{m_1+m_2}{\omega_1+\omega_2}\right),
b_2=\frac{\omega_1+\omega_2}{m_1+m_2}, b_3=-\frac{(m_1-m_2)}{m_1\omega_2+m_2\omega_1},
b_4=\frac{(\omega_1+\omega_2)}{(m_1\omega_2+m_2\omega_1)}.$$

\subsection{For vector meson with quantum numbers
$J^{P}=1^{-}$}

The general form for the relativistic wavefunctions of vector
state $J^P=1^-$(or $J^{PC}=1^{--}$ for quarkonium) can be written
as eight terms, which are constructed by $P_{f1}$, $q_{f1\perp}$, $\epsilon_1$ and gamma matrices~\cite{glwang},
\begin{eqnarray}
\varphi_{1^{-}}(\vec q_{f1})&=&
q_{f1\perp}\cdot{\epsilon}_1
\left[f'_1+\frac{\not\!P_{f1}}{M_{f1}}f'_2+
\frac{{\not\!q}_{f1\perp}}{M_{f1}}f'_3+\frac{{\not\!P_{f1}}
{\not\!q}_{f1\perp}}{M_{f1}^2} f'_4\right]+M_{f1}{\not\!\epsilon_1}f'_5\\ \nonumber
&+&
{\not\!\epsilon_1}{\not\!P_{f1}}f'_6+
({\not\!q}_{f1\perp}{\not\!\epsilon_1}-
q_{f1\perp}\cdot{\epsilon_1})
f'_7+\frac{1}{M_{f1}}({\not\!P_{f1}}{\not\!\epsilon_1}
{\not\!q}_{f1\perp}-{\not\!P_{f1}}q_{f1\perp}\cdot{\epsilon_1})
f'_8,\label{eq13}
\end{eqnarray}
where ${\epsilon}_1$ is the polarization vector of the
vector meson in the final state.

Due to
the last two equations of Eq.~(\ref{eq11}):
$\varphi_{0^-}^{+-}=\varphi_{0^-}^{-+}=0$, we have~\cite{pwave4}:
$$f'_1=\frac{\left[q_{f1\perp}^2 f'_3+M_{f1}^2f'_5
\right] (m'_1m'_2-w'_1w'_2+q_{f1\perp}^2)}
{M_{f1}(m'_1+m'_2)q_{f1\perp}^2},~~~f'_7=\frac{f'_5M_{f1}(-w'_1+w'_2)}
{(m'_1w'_2+m'_2w'_1)},$$
$$f'_2=\frac{\left[-q_{f1\perp}^2 f'_4+M_{f1}^2f'_6\right]
(m'_1w'_2-m'_2w'_1)}
{M_{f1}(w'_1+w'_2)q_{f1\perp}^2},~~~f'_8=\frac{f'_6M_{f1}(w'_1w'_2-m'_1m'_2-q_{f1\perp}^2)}
{(m'_1+m'_2)q_{f1\perp}^2}.$$

The relativistic positive wavefunctions of $^3S_1$ state can be written as~\cite{Bs}:
\begin{eqnarray}
{\varphi}_{1^{-}}^{++}(\vec{q}_{f1})&=&b_1\not\!{\epsilon}_1+b_2\not\!{\epsilon}_1\not\!{P_{f1}}
+b_3(\not\!{q_{f1\bot}}\not\!{\epsilon}_1-q_{f1\bot}\cdot{\epsilon}_1)
+b_4(\not\!{P_{f1}}\not\!{\epsilon}_1\not\!{q_{f1\bot}}-\not\!{P_{f1}}q_{f1\bot}\cdot{\epsilon}_1)
\nonumber\\
&&+q_{f1\bot}\cdot{\epsilon}_1(b_5+b_6\not\!{P_{f1}}+b_7\not\!{q_{f1\bot}}+b_8\not\!{q_{f1\bot}}\not\!{P_{f1}}),
\end{eqnarray}
where we first define the parameter $n_i$ which are the functions of
$f'_i$ ($^3S_1$ wave functions):
$$n_1=f'_5
-f'_6\frac{(w'_1+w'_2)}{(m'_1+m'_2)}, n_2=f'_5
-f'_6\frac{(m'_1+m'_2)}{(w'_1+w'_2)}, n_3=f'_3
+f'_4\frac{(m'_1+m'_2)}{(w'_1+w'_2)},$$ then we define
the parameters $b_i$ which are the functions of $f'_i$ and $n_i$:
$$b_1=\frac{M_{f1}}{2}n_1, b_2=-\frac{(m'_1+m'_2)}{2(w'_1+w'_2)}n_1,
 b_3=\frac{M_{f1}(w'_2-w'_1)}{2(m'_1w'_2+m'_2w'_1)}n_1, b_4=\frac{(w'_1+w'_2)}{2(w'_1w'_2+m'_1m'_2-{q_{f1\bot}^{2}})}n_1,$$
 $$b_5=\frac{1}{2M_{f1}}\frac{(m'_1+m'_2)(M_{f1}^2n_2+{q_{f1\bot}^{2}}n_3)}{(w'_1w'_2+m'_1m'_2+{q_{f1\bot}^{2}})},
  b_6=\frac{1}{2M_{f1}^2}\frac{(w'_1-w'_2)(M_{f1}^2n_2+{q_{f1\bot}^{2}}n_3)}{(w'_1w'_2+m'_1m'_2+{q_{f1\bot}^{2}})},$$
$$b_7=\frac{n_3}{2M_{f1}}-\frac{f'_6M_{f1}}{(m'_1w'_2+m'_2w'_1)},
 b_8=\frac{1}{2M_{f1}^2}\frac{w'_1+w'_2}{m'_1+m'_2}n_3-f'_5\frac{w'_1+w'_2}{(m'_1+m'_2)(w'_1w'_2+m'_1m'_2-{q_{f1\bot}^{2}})}.$$

\section{The formulation of two-body open charm strong decays}

For the two-body OZI-allowed open charm strong decays, such as $\eta_c(3S)\to D\bar D^*$,
we adopt the $^3P_0$ model to calculate the strong decay amplitude.
The non-relativistic $^3P_0$ model describe the decay matrix elements by the $q\bar q$
pair-production Hamiltonian: $H=g\int d^3x\bar\psi\psi$~\cite{3p04}.
According to the improved B-S method which is a relativistic model,
we can extend the  non-relativistic $^3P_0$ model to the relativistic
form:  $H=-ig\int d^4x\bar\psi\psi$~\cite{th4,3p06}.
Here $\psi$ is the dirac quark field, $g=2m_q\gamma$,
$m_q$ is the quark mass of the light quark-pairs,
$\gamma$ is a dimensionless constant which describe the pair-production strength
and can be obtained by fitting the experimental data.
In this paper, we choose $\gamma=0.483$~\cite{3p05} which give reasonable calculation of $\eta_c(3S)$,
then we use the same value to $\eta_c(4S)$.

\begin{figure}[htbp]
\centering
\includegraphics[height=5cm]{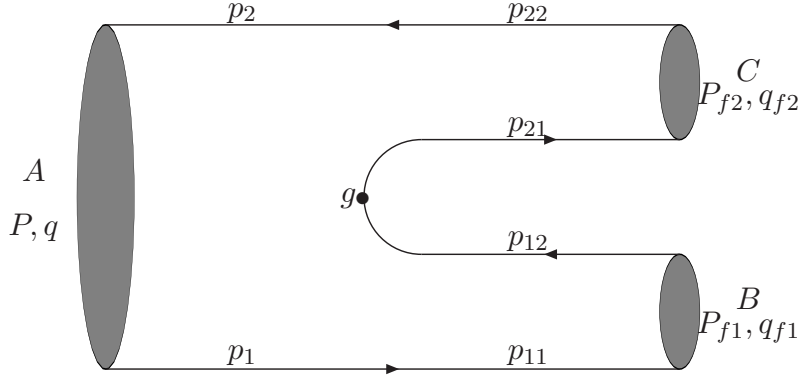}
\caption{\label{OZIStrongdecay}The feynman diagram of two-body open charm strong decay.}
\end{figure}

Using the $q\bar q$ pair-production Hamiltonian,
the amplitude of two-body OZI-allowed open charm strong decays $A\to B+C$ in Fig.~\ref{OZIStrongdecay},
can be written as~\cite{th4,3p06},
\begin{eqnarray}\label{amplitude}
<BC|H|&A>&=-ig\int\frac{d^4q}{(2\pi)^4}
{\rm Tr}[\chi_P(q)S_2^{-1}(p_2)
\bar\chi_{P_{f2}}(q_{f2})\bar\chi_{P_{f1}}(q_{f1})S_1^{-1}(p_1)]\nonumber\\
&=&g\int\frac{d^3\vec q}{(2\pi)^3}{\rm Tr}\left[\frac{\not\!{P}}{M}\varphi^{++}_P(\vec q)
\frac{\not\!{P}}{M}\bar\varphi^{++}_{P_{f2}}(\vec q_{f2})\bar\varphi^{++}_{P_{f1}}(\vec q_{f1})\right]
\left(1-\frac{M-w_1-w_2}{2w_{12}}\right),
\end{eqnarray}
where $\varphi^{++}_P(\vec q)$, $\varphi^{++}_{P_{f1}}(\vec q_{f1})$ and $\varphi^{++}_{P_{f2}}(\vec q_{f2})$
are the relativistic positive wavefunctions of initial meson $A$, finial meson $B$ and $C$, respectively.
$\bar\varphi=\gamma^0\varphi^\dagger\gamma^0$.
We have given the detailed form of wavefunctions in Sec.~III.
$P$, $P_{f1}$, $P_{f2}$ and $\vec q$, $\vec q_{f1}$, $\vec q_{f2}$
are the momentum and three dimension relative momentum between quark and anti-quark of initial meson $A$,
finial meson $B$ and $C$, respectively.
$\vec{q}_{f1}=\vec q-\frac{m_c}{m_c+m_{u,d,s}}\vec{P}_{f1}$,
$\vec{q}_{f2}=\vec q+\frac{m_c}{m_c+m_{u,d,s}}\vec{P}_{f2}$.
$\vec{P}_{f1}$ and $\vec{P}_{f2}$ are the three momentum of finial mesons $B$ and $C$.
$w_{12}=\sqrt{m_{u,d,s}^2+\vec q^2_{f1}}$.

Using the B-S wavefunctions in Sec.~III and the formula of amplitude Eq.~(\ref{amplitude}),
the two-body open charm strong decay amplitude can be defined as,
\begin{eqnarray}\label{amplitude1}
&&\mathcal M(A\to D\bar D^*)=\epsilon_{1\mu}P^\mu t_1,\nonumber\\
&&\mathcal M(A\to D^*\bar D^*)=\varepsilon_{\mu\nu\alpha\beta}P^\mu P^\nu_{f1}\epsilon^{\alpha}_1\epsilon^{\beta}_2 t_2,
\end{eqnarray}
where $A$ denote $\eta_c(3S)$ or $\eta_c(4S)$,
$\epsilon_{1}$ and $\epsilon_{2}$ are the polarization vector of the final mesons $B$ and $C$.
$t_1$ and $t_2$ are the strong decay coupling constants which are related to the B-S wavefunctions.

Finally, using Eq.~(\ref{amplitude1}) the two-body open charm strong decay width can be written as,
\begin{eqnarray}\label{width}
\Gamma=\frac{|\vec P_{f1}|}{8\pi M^2}\sum_{\lambda}|\mathcal M|^2,
\end{eqnarray}
where $|\vec P_{f1}|=\sqrt{[M^2-(M_{f1}-M_{f2})^2][M^2-(M_{f1}+M_{f2})^2]}/(2M)$
which is the three momentum of the final mesons.

\section{Number results and discussions}

In order to fix Cornell potential in Eq.(\ref{eq16}) and masses of quarks,
 we take these parameters: $a=e=2.7183,
\lambda=0.210$ GeV$^2$, ${\Lambda}_{QCD}=0.270$ GeV, $\alpha=0.060$
GeV, $m_u=0.305$ GeV, $m_d$=0.311 GeV, $m_s$=0.500 GeV, $m_b=4.96$ GeV, $m_c=1.62$ GeV, $etc$~\cite{mass1},
which are best to fit the mass spectra of ground states $B$, $D$ mesons and other heavy mesons.
And we get the masses: $M_{D^\pm}=1.869$ GeV,
$M_{D^0}=1.865$ GeV, $M_{D_s^\pm}=1.968$ GeV, $M_{D^{*0}}=2.007$ GeV,
$M_{D^{*\pm}}=2.010$ GeV, $M_{D_s^{*\pm}}=2.112$ GeV,
$M_{\eta_c(3S)}$=3.942 GeV, $M_{\eta_c(4S)}$=4.156 GeV.

\begin{table}[htbp]
\caption{\label{exclusivewidth}The exclusive strong decay widths of $\eta_c(3S)$ and $\eta_c(4S)$ (unit in MeV).}
\begin{center}
\begin{tabular}{|c|ccc|c|ccc|}
\hline \hline
Mode&$D^0\bar D^{*0}$&$D^+D^{*-}$&$D\bar D^*$&$D^-_sD^{*+}_s$&$D^{*0}\bar D^{*0}$&$D^{*-} D^{*+}$&$D^*\bar D^*$  \\ \hline
$\eta_c(3S)$& $18.0^{+9.0}_{-7.7}$& $15.5^{+9.4}_{-7.6}$&$33.5^{+18.4}_{-15.3}$&--&--&--&--\\ \hline
$\eta_c(4S)$&$27.7^{+4.7}_{-5.6}$&$27.0^{+5.7}_{-6.0}$&$54.7^{+10.4}_{-11.6}$&$0.28^{+0.24}_{-0.15}$&$7.7^{+6.1}_{-4.9}$&$7.2^{+5.7}_{-4.4}$&$14.9^{+11.8}_{-9.3}$\\
\hline \hline
\end{tabular}
\end{center}
\end{table}

Considering $X(3940)$ as $\eta_c(3S)$ state,
there is only one decay mode: $0^-\to 1^-0^-$,
According to the kinematic ranges,
the corresponding final states are: $D^{0}\bar D^{*0}$, $\bar D^{0} D^{*0}$,
$D^{+}D^{*-}$ and $D^{*+}D^{-}$.
Considering $X(4160)$ as $\eta_c(4S)$ state,
there are two decay mode: $0^-\to 1^-0^-$ and $0^-\to 1^-1^-$,
within the kinematic ranges,
the corresponding decay channels include: $D^{0}\bar D^{*0}$, $\bar D^{0} D^{*0}$,
$D^{+}D^{*-}$, $D^{*+}D^{-}$, $D^+_sD^{*-}_s$, $D^-_sD^{*+}_s$, $D^{*0}\bar D^{*0}$ and $D^{*-} D^{*+}$.
We have shown the exclusive two-body open charm strong decay widths of $\eta_c(3S)$ and $\eta_c(4S)$ in Table.~\ref{exclusivewidth},
where $D\bar D^*$ means $D^0\bar D^{*0}$+$D^+D^{*-}$,
and $D^*\bar D^*$ means $D^{*0}\bar D^{*0}$+$D^{*-} D^{*+}$.
for $D^0\bar D^{*0}$, $D^+D^{*-}$ and $D^-_sD^{*+}_s$,
we have considered the isospin conservation of the final mesons.
In Table.~\ref{totalwidth}, we have presented the total widths with different theoretical model and the experimental data for convenience.
We also consider the uncertainties by varying all the input parameters
simultaneously within $\pm$5$\%$ of the central values in Table.~\ref{exclusivewidth} and Table.~\ref{totalwidth}.

\begin{figure}[htbp]
\centering
\includegraphics[height=6cm]{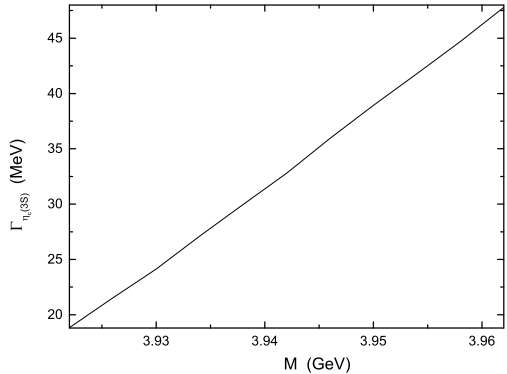}
\caption{\label{X3940m-width}The relation of decay width to the mass of $\eta_c(3S)$.}
\end{figure}

In Table.~\ref{exclusivewidth},
we find that the dominant strong decay channels of $\eta_c(3S)$ is $D\bar D^*$,
and agree with the experimental observation by Belle collaboration~\cite{3940,39404160}.
The total two-body open charm strong decay widths of $\eta_c(3S)$ is $\Gamma_{\eta_c(3S)}=(33.5^{+18.4}_{-15.3})$ MeV,
which is smaller than the result of Ref.~\cite{Yan1},
but it is in accordance with experimental results.
So $\eta_c(3S)$ could be a good candidate of the $X(3940)$.
Because of the mass of $X(3940)$ has the errors,
we plot the relations of decay widths of $\eta_c(3S)$ to
the masses of $\eta_c(3S)$ in Fig.~\ref{X3940m-width},
the relations of decay widths to the masses of $\eta_c(3S)$ are linear.
The decay widths increase with the increase of the masses of $\eta_c(3S)$.

\begin{table}[htbp]
\caption{\label{totalwidth}The total strong decay widths of $\eta_c(3S)$ and $\eta_c(4S)$ (unit in MeV).
`Ex.' means the experimental data of $X(3940)$ and $X(4160)$ from
PDG~\cite{PDG}.}
\begin{center}
\begin{tabular}{|ccccc|}
\hline \hline
Mode&Ours&\cite{41601}&\cite{Yan1}&Ex  \\ \hline
$\Gamma_{\eta_c(3S)}$& $33.5^{+18.4}_{-15.3}$&--&$99.8\pm12.0$&$37^{+26}_{-15}\pm8$ \\ \hline
$\Gamma_{\eta_c(4S)}$&$69.9^{+22.4}_{-21.1}$&25.0&--&$139^{+111}_{-61}\pm21$\\
\hline \hline
\end{tabular}
\end{center}
\end{table}

\begin{figure}[htbp]
\centering
\includegraphics[height=5cm]{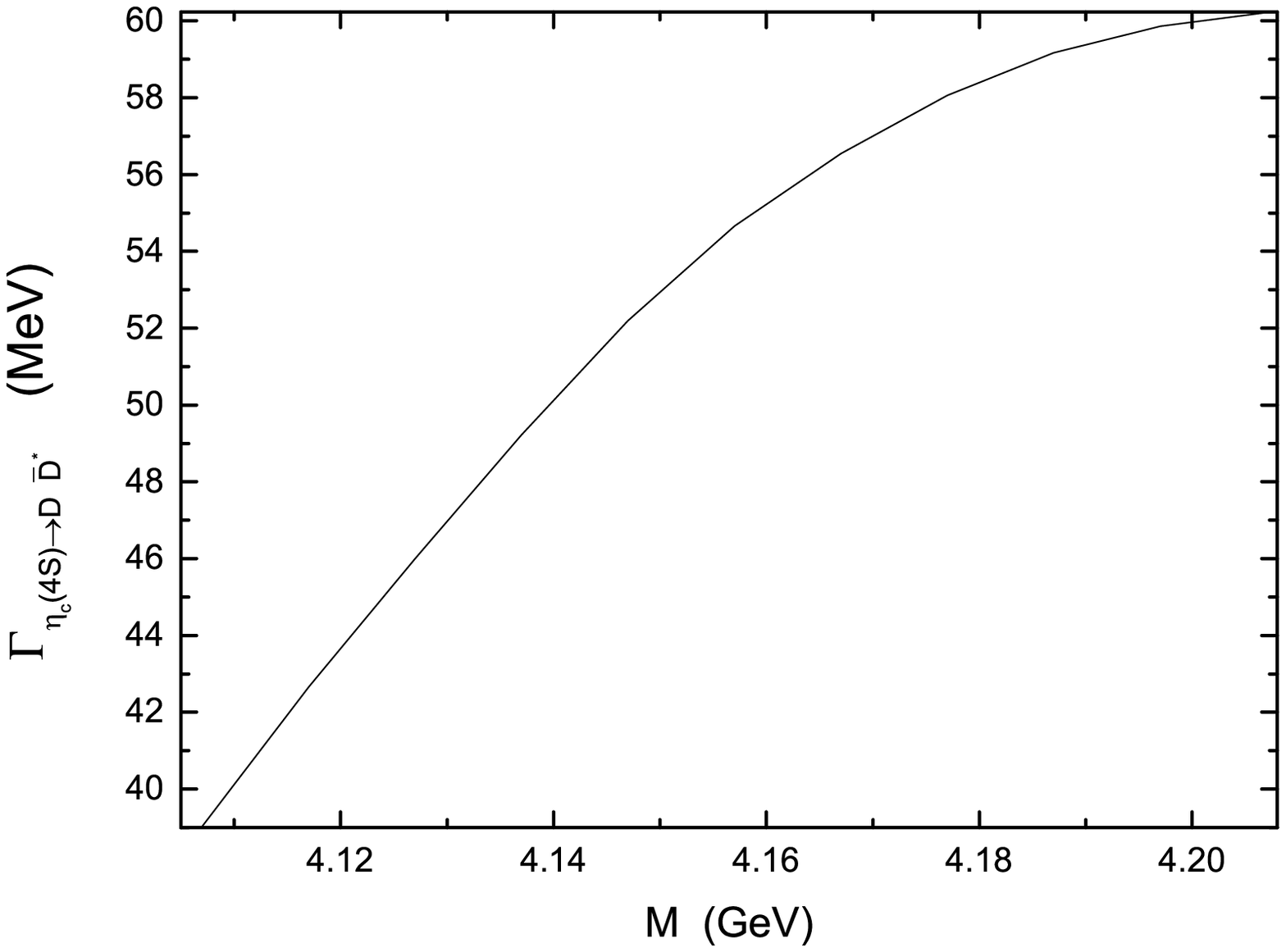}
\includegraphics[height=5cm]{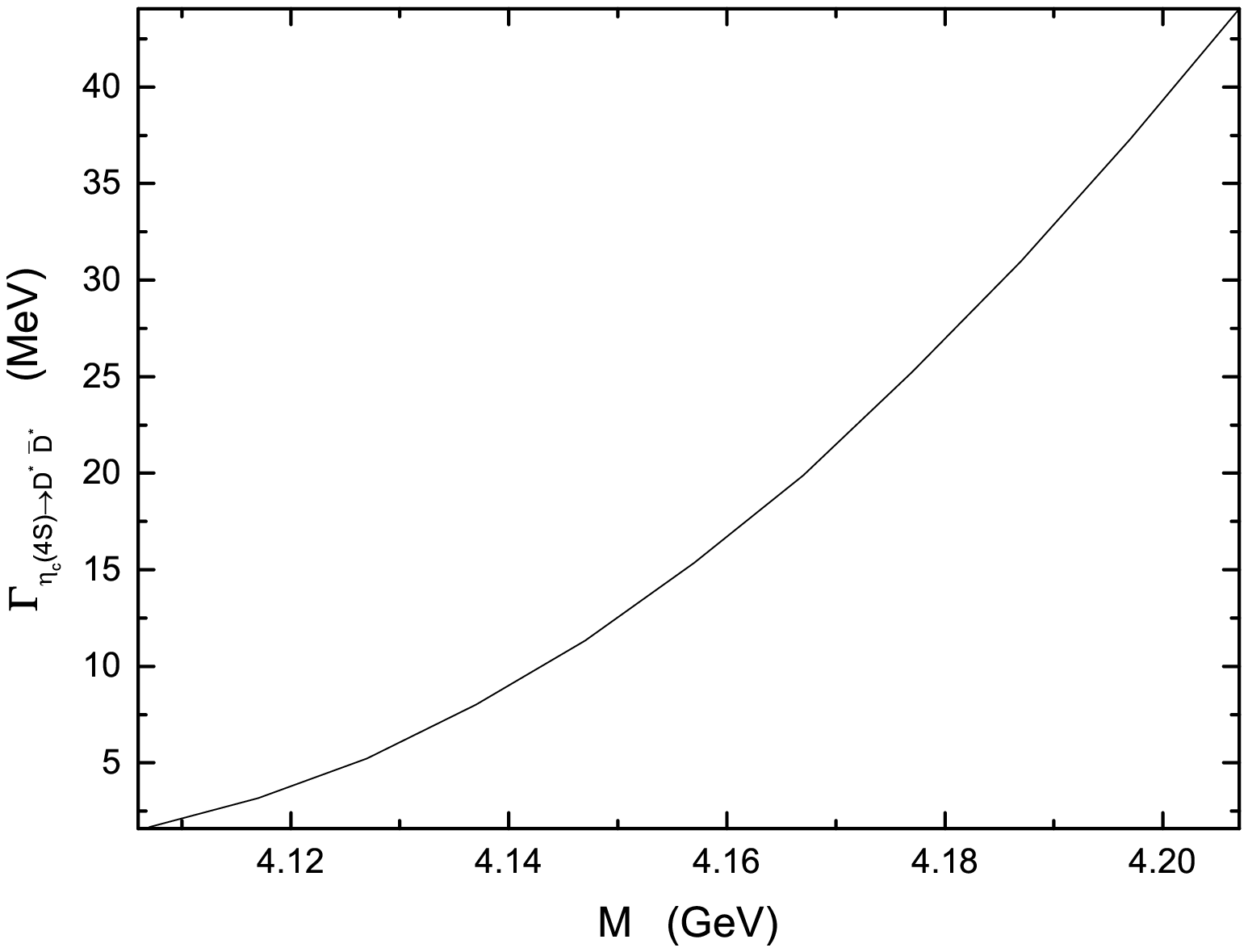}
\caption{\label{X4160m-width}The relation of different decay width to the mass of $\eta_c(4S)$.}
\end{figure}

\begin{figure}[htbp]
\centering
\includegraphics[height=6cm]{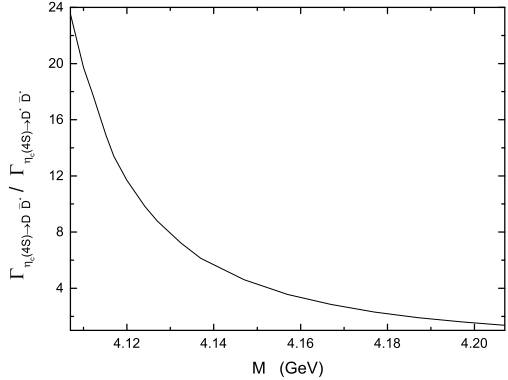}
\caption{\label{Br}The relation of $\Gamma_{\eta_c(4S)\to D\bar D^*}/\Gamma_{\eta_c(4S)\to D^*\bar D^*}$ to the mass of $\eta_c(4S)$.}
\end{figure}

For $\eta_c(4S)$ state, the main strong decay channels are $D\bar D^*$ and $D^*\bar D^*$,
$\eta_c(4S)\to D^-_sD^{*+}_s$ is very small with the small phase space,
and the decay $\eta_c(4S)\to D\bar D$ is forbidden.
In Table.~\ref{totalwidth},
the total two-body open charm strong decay widths of $\eta_c(4S)$ is $\Gamma_{\eta_c(4S)}=(69.9^{+22.4}_{-21.1})$ MeV.
Our result is larger than the result of Ref.~\cite{41601},
but considering the uncertainties of the results, our result is closed to the lower limit of $X(4160)$ for experimental data~\cite{39404160}.
In our calculation, the ratio of the decay width
$\frac{\Gamma(\eta_c(4S)\to D\bar D)}{\Gamma(\eta_c(4S)\to D^*\bar D^*)}=0$,
which is consistent with the experimental data
$\frac{\Gamma(X(4160)\to D\bar D)}{\Gamma(X(4160)\to D^*\bar D^*)}<0.09$~\cite{39404160}.
There is another ratio of the decay width:
$\frac{\Gamma(\eta_c(4S)\to D\bar D^*)}{\Gamma(\eta_c(4S)\to D^*\bar D^*)}=3.67$,
which is much larger than the upper limit of the experimental data
$\frac{\Gamma(X(4160)\to D\bar D^*)}{\Gamma(X(4160)\to D^*\bar D^*)}<0.22$ which is reported by Belle~\cite{39404160}.
In order to find out the relation of the decay width to the mass of $\eta_c(4S)$,
we plot the relation of different decay width and decay ratio to the mass of $\eta_c(4S)$ in Fig.~\ref{X4160m-width}
and Fig.~\ref{Br}.
Especially in Fig.~\ref{Br}, the decay ratio is decreased with the increased mass of $\eta_c(4S)$,
but the decay ratio is larger than the experimental data at large mass,
so $\eta_c(4S)$ is not the candidate of $X(4160)$,
and more investigations of $X(4160)$ is needed in future.

In summary,
considering $X(3940)$ and $X(4160)$ as $\eta_c(3S)$ and $\eta_c(4S)$ states,
we study the two-body open charm OZI-allowed strong decay of $\eta_c(3S)$ and $\eta_c(4S)$ by the improved B-S
method combine with the $^3P_0$ model.
For the strong decay of $\eta_c(3S)$, the dominant strong decay is $\eta_c(3S)\to D\bar D^*$,
the corresponding strong decay width is $\Gamma_{\eta_c(3S)}=(33.5^{+18.4}_{-15.3})$ MeV,
which is closed to the experimental data,
therefore, $\eta_c(3S)$ is a good candidate of $X(3940)$.
For $\eta_c(4S)$ state, the main strong decay channels are $D\bar D^*$ and $D^*\bar D^*$,
$\eta_c(4S)$ can not decay to $D\bar D$,
which have not been observed for $X(4160)$ in experiment.
$\Gamma (D^*\bar D^*)$ is smaller than $\Gamma(D\bar D^*)$,
the ratio of the decay width $\frac{\Gamma(D\bar D^*)}{\Gamma (D^*\bar D^*)}$ is larger than
the experimental data by Belle.
We also find that  the ratio of the decay width $\frac{\Gamma(D\bar D^*)}{\Gamma (D^*\bar D^*)}$ is
dependent on the mass of $\eta_c(4S)$.
Finally, we calculate the strong decay width of $\eta_c(4S)$:
$\Gamma_{\eta_c(4S)}=(69.9^{+22.4}_{-21.1})$ MeV,
considering the errors of the results,
it's closed to the lower limit of $X(4160)$.
With large errors of full decay width, it's hard to confirm that $\eta_c(4S)$ is the candidate of $X(4160)$.
But the ratio of the decay width $\frac{\Gamma(D\bar D^*)}{\Gamma (D^*\bar D^*)}$ is not consistent with the experimental data,
so taking the $\eta_c(4S)$ as an assignment of $X(4160)$ can be excluded
and more investigations is needed in future.

\noindent
{\Large \bf Acknowledgements}
This work was supported in part by
the National Natural Science Foundation of China (NSFC) under
Grant No.~11405004, No.~11405037, No.~11505039, No.~11575048
and the Science and technology research project of Ningxia high school No.~NGY2015142.

\end{document}